\input epsf.def
\documentstyle[prl,twocolumn,aps]{revtex}
\begin{document}

\twocolumn[\hsize\textwidth\columnwidth\hsize\csname @twocolumnfalse\endcsname
\draft
\tolerance 50000

\title{Correlated fermions in a one-dimensional quasiperiodic potential}
\author{Julien Vidal$^{1}$,  Dominique Mouhanna$^{2}$ and Thierry
Giamarchi$^{3}$}
\address{$^1$ Groupe de Physique des Solides, CNRS UMR 7588,
Universit\'{e}s Pierre et Marie Curie Paris 6 et Denis Diderot Paris 7,\\
2 place Jussieu, 75251 Paris Cedex 05 France}
\address{$^2$ Laboratoire de Physique Th\'{e}orique et
Hautes Energies, CNRS UMR 7589, Universit\'{e} Pierre et Marie Curie Paris
6, \\
4 place Jussieu, 75252 Paris Cedex 05 France}
\address{$^3$ Laboratoire de physique des Solides, CNRS UMR 8502,
UPS B{\^a}t 510, 91405 Orsay, France}
\maketitle

\begin{abstract}
We study analytically one-dimensional interacting spinless
fermions in a Fibonacci potential. We show that the effects of the
quasiperiodic modulation are intermediate between those of a
commensurate potential and a disordered one. The system exhibits a
metal-insulator transition whose position depends both on the
strength of the correlations and on the position of the Fermi level.
Consequently, the conductivity displays a power law like
size and frequency behaviour characterized by a non trivial
exponent.
\end{abstract}

\vskip2pc]

Since the discovery of quasicrystals (QCs) in 1984 by Schechtman
{\it et al.} \cite{Schechtman}, the electronic properties of
quasiperiodic systems have been intensively studied. These
metallic alloys are notably characterized by a low electrical
conductivity $\sigma$ which increases when either temperature or
disorder increases \cite{Mayou93}. The very low temperature
behaviour of $\sigma$ is still an open question and
depends on the materials. For example, in AlCuFe \cite{Klein} and
AlCuRu \cite{Biggs}, a finite conductivity at zero temperature is
expected whereas recent results \cite{Delahaye} seem to confirm a
Mott's variable range hopping mechanism ($\sigma(T)\sim
\exp-(T_0/T)^{1/4}$) for $i$-AlPdRe icosahedral phase down to $20$
mK. The optical conductivity is also unusual since there is no
Drude Peak for icosahedral quasicrystals \cite{Homes,Burkov}.

Many theoretical works have attempted to understand how the
quasiperiodic order could induce such exotic behaviours. In particular,
the case of independent electrons in one-dimensional ($1D$) systems has
been deeply investigated for different structures (Harper model,
Fibonacci chain,..)\cite{KKT,Ostlund}, giving rise to singular
continuous spectra with an infinite number of gaps. Moreover, the
corresponding eigenstates are neither extended nor localized but
critical, and are known to be responsible of anomalous diffusion
\cite{Zhong,Piechon}. For higher dimensional systems (Penrose tiling,
Octagonal tiling, Icosahedral structure...), similar studies had also
displayed complex and intricated spectra, with analogous characteristics
of the electronic states
\cite{Passaro_Octo_diffusion,Yamamoto_Penrose}. The transport
properties have also been widely studied
\cite{Sire_Aussois,Roche_Review,Kubo_Goda,Roche_Fermi,Bellissard_aperiodic}
within different approach (Landauer conductance, Bloch-Boltzmann
approximation, Kubo-Greenwood conductivity,...). However, given the complexity
of these problems due to the geometry alone, the interactions between
electrons have often been neglected. Even in $1D$ incommensurate
structures, few results have been obtained
\cite{Hiramoto_HF,Chaves_meso,Chaves_HF,Mastro,Sen}.

In this Letter, we investigate the effect of the interactions considering
a Hubbard-like model for spinless fermions embedded in a Fibonacci
potential. We take  the correlations into
account using a bosonization technique whereas the
quasiperiodicity is treated perturbatively.
Using a renormalization group approach,
we show that the quasicrystalline system displays a metal-insulator
transition (MIT) induced by the interactions. The corresponding critical
MIT point is found to be strongly dependent on the Fermi level.
In marked contrast with the simple cases of
disordered or commensurate potentials, the quasiperiodicity
leads to a power law like dependence of the conductivity
either in size or frequency with an exponent depending both on
the interactions and on the position of the Fermi level.
Though our analysis is performed for a Fibonacci potential, we stress
that these results can be extended to any potential having a non flat dense
Fourier spectrum.

\begin{figure}
\epsfscale=400
\centerline{\epsfbox{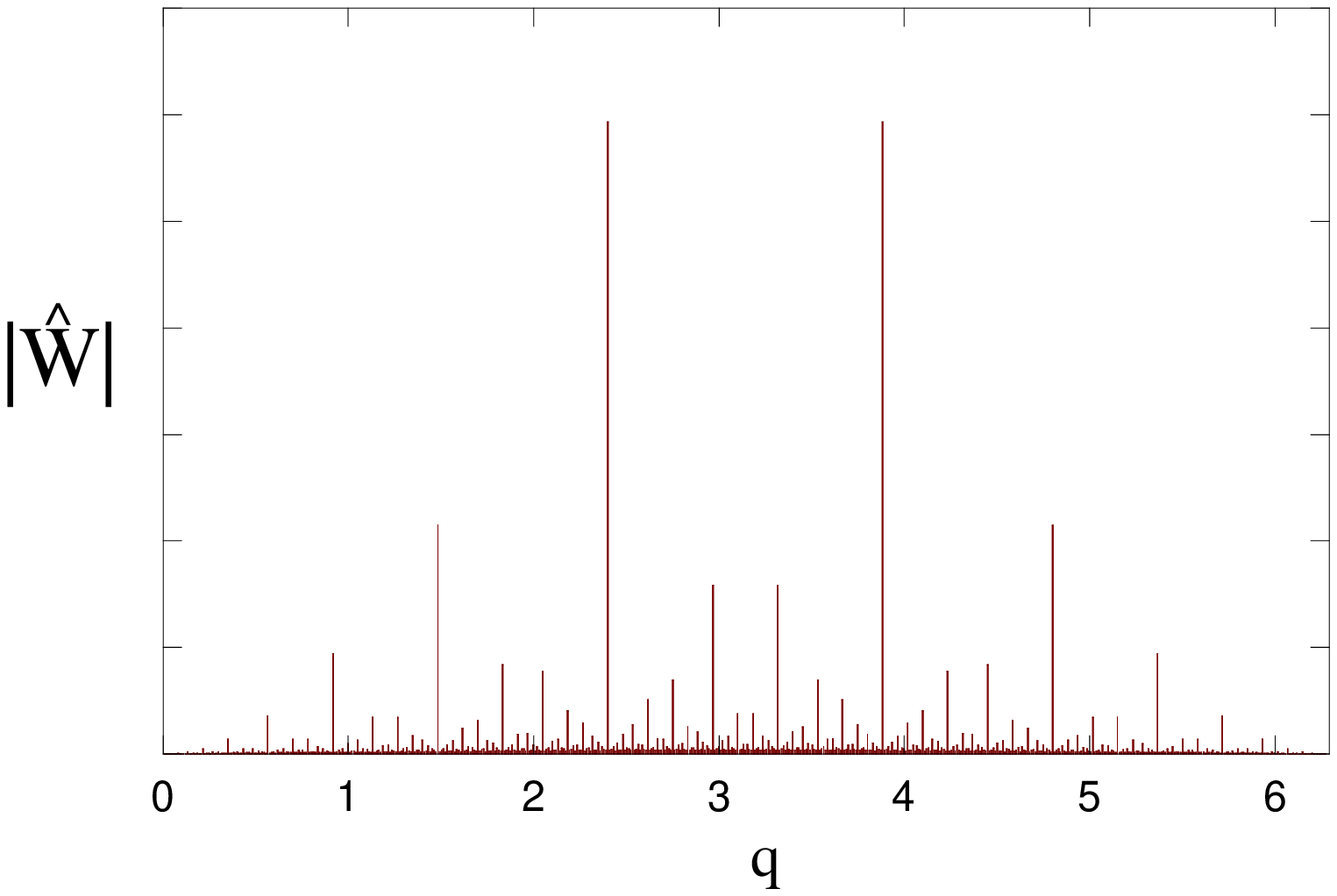}}
\caption{\label{TF_Fibo}
Fourier transform of the diagonal Fibonacci potential for the $15^{th}$
approximant with $F_{15}=610$ sites per cell.}
\end{figure}

Let us consider a model of interacting spinless fermions on a
quasiperiodic lattice described by the following Hamiltonian:
\begin{equation}
H=-t \sum_{\langle i,j \rangle} c^\dagger_{i}\,c_{j}+V \sum_{\langle i,j
\rangle} n_i\,n_j
+ \sum_{i} W_i\,n_i
\label{hamil}
\end{equation}
where $c^\dagger_{i}$ (resp. $c_{i}$) denotes the creation (resp.
annihilation) fermion operator,
 $n_i=c^\dagger_{i}\,c_{i}$ represents the fermion density on site $i$, and
$\langle \ldots \rangle$ stands for
nearest neighbors pairs. The quasiperiodicity is provided by the $W_i$'s
that take
two discrete values $W_A=+\lambda/2$ or $W_B=-\lambda/2$
given by the spatial modulation of the Fibonacci chain. In
fact, we consider a periodic approximant
of this structure with $F_l$ sites per unit cell that can be obtained
 by $l$ iterations of the substitution rules: $A\rightarrow
AB,\hspace{2.ex} B\rightarrow A$,
where $F_l$ is the $l^{th}$ element of the Fibonacci sequence defined by:
\begin{eqnarray}
F_1&=&F_2=1 \nonumber\\
F_{l+1}&=&F_{l}+F_{l-1}
\label{recur}
\end{eqnarray}
We denote $p=F_{l-2}, s=F_{l-1}, n=F_{l}$ and $n'=s$ (resp. $n'=p$) if
$l$ is even (resp. odd). In the quasiperiodic limit ($l\rightarrow
\infty$), the ratio $s/p$ converges toward the golden mean
$\tau={1+\sqrt{5} \over 2}$.
It is also useful to compute the Fourier transform of the potential $W$. This
can be done {\it via} the conumbering scheme
\cite{Mosseri_conumbering}:
\begin{equation}
\hat W\left(q={2\pi m\over na}\right)={\lambda \, e^ {i{\pi m  n' (s-1)
\over n}} \sin
\left({\pi m n' s \over n}\right)
\over n \sin \left({\pi m n'\over n}\right)}
\label{TF}
\end{equation}
for $m=1$ to $n-1$ ($a$ is the lattice spacing). A global shift of the $W_i$
allows us to deal with a zero-averaged potential
so that we can set $\hat W(0)=0$. The interest of this potential lies in
the fact
that it is perturbatively non trivial (the Fourier
transform is dense in $[0, 2\pi]$, in the quasiperiodic limit).
Moreover, the underlying substitution rule provides a self-similar structure
that can be readily seen in Fig.~\ref{TF_Fibo}.
If the quasiperiodic perturbation is small enough, we can focus on the
low energy properties around the Fermi level.
To treat the interactions, it is convenient to use a
bosonic representation of the fermion operators.
In this representation the hamiltonian (\ref{hamil}) becomes:
\begin{equation} \label{Hbos}
H=H_0+H_W
\end{equation}
where $H_0$ stems from the periodic part ($\lambda=0$) of (\ref{hamil}) and
reads in the continuum limit \cite{schulz_houches_revue}:
\begin{equation} \label{H0bos}
H_0={1\over 2\pi}\int dx \left[(u K)(\pi \Pi)^2+\left({u\over K}\right)
(\partial_x \phi)^2\right].
\end{equation}
In (\ref{H0bos}) $\phi$ is a boson field related to the long wave length
part of the
fermionic density by $\rho(x)=-\nabla \phi(x)/\pi$, and $\Pi$ is its
canonically conjugate field. All the interactions are absorbed in the two
constants $u$ and $K$, where $u$ is the renormalized Fermi velocity (in the
non-interacting case ($K=1$), $u=v_F=2ta\sin(k_F a)$), and $K$ is the
parameter controlling the decay of various correlation functions. For
weak interactions, $u$ and $K$ can be perturbatively expressed in terms
of the microscopic parameters $t$ and $V$:
\begin{equation}\label{param}
uK=v_F \qquad,\qquad
\displaystyle{u\over K}=v_F+{2V\over \pi}
\end{equation}
Actually, the representation (\ref{H0bos}) is more general
\cite{haldane_bosonisation,haldane_xxzchain} and gives the correct low
energy description of the system, even when the interactions are strong
provided the exact $u$ and $K$ parameters are used. The quasiperiodic
part $H_W$ can also be written in terms of the boson fields
\cite{caution_shift}:
\begin{equation}
H_W = \frac1{2\pi \alpha} \int dx \,W(x)  \cos\left[2k_Fx+2\phi(x)\right]
\label{hquasi}
\end{equation}
where $k_F$ is the Fermi wave vector and $\alpha$ is a short distance
cut-off of the order of the lattice constant $a$.

The effect of the quasiperiodic potential is computed using a
perturbative renormalization group (RG) approach, similar to the one
for a single harmonic \cite{giamarchi_mott_ref}. The RG equations for
the potential and the interaction parameter read:
\begin{eqnarray}
{dK\over dl}&=&-K^2 G(l) \label{recy1}\\
{dy_q\over dl}&=&(2-K) \, y_q\label{recy2} \\
G(l) &=& \sum_{\varepsilon=\pm 1}\sum_q y_q^2 R\left[(q+\varepsilon
2k_F)\,\alpha(l) \right] \label{recy3}
\end{eqnarray}
where $y_q=\alpha \hat W(q) / u$ is the dimensionless Fourier components of
$W$,
$\alpha(l)=\alpha(0) \,e^l$ is the renormalized short distance cutoff.
In (\ref{recy3}), the sum over $q$ is performed for
$q=2\pi m/n$ with $m\in [1,n-1]$, and  $R$ is an ultraviolet regulator
whose precise form depends on the cut-off procedure used.
Without loss of generality, we choose a gaussian regulator $R(x)=e^{-x^2}$.
Note that the renormalization  of the velocity $u$ is neglected since
it is of higher order in the potential amplitude $\lambda$.

The physical properties of the system are determined by
the long distance behaviour of $K$. If $K$ converges toward a fixed
point, the system remains metallic but with renormalized parameter
$(K^*,u^*)$. Otherwise, the solution flows to a strong coupling regime
whose physics depends on the precise nature of the potential $W$.
Note that expression (\ref{hquasi}) is  valid for any lattice
potential perturbatively treated. For a single harmonic ($\hat{W}(q)=\lambda \,
\delta_{q,q_0}$) two situations occurs. If $q_0\ne 2k_F$, R stops
the renormalization of $K$ at large enough lengthscale
and the potential is irrelevant. {\it A contrario},
if $q_0=2k_F$ one recovers the usual Metal-Insulator transition at $K_c=2$.
In the disordered case, the $\hat W(q)$'s are
given by an uniform averaged  distribution:
$\overline{\hat W^*(q)\hat W(q')}=y\,\delta_{qq'}$.
In the limit of weak disorder, (\ref{recy2}) can be integrated
neglecting the renormalization of $K$: $y(l)=y(0)\,e^{(2-K)l}$.
Then, Eq.(\ref{recy1}) simply becomes:
\begin{equation} \label{eq:mu}
\frac{dK}{dl} = - K^2 C \,e^{(3-2K)l}
\end{equation}
where $C$ is a constant.
Eq.(\ref{eq:mu}) defines a critical value $K_c=3/2$, separating an
insulating phase ($K<K_c$) from a metallic state ($K>K_c$)
\cite{giamarchi_loc}.

\begin{figure}
\epsfscale=400
\centerline{\epsfbox{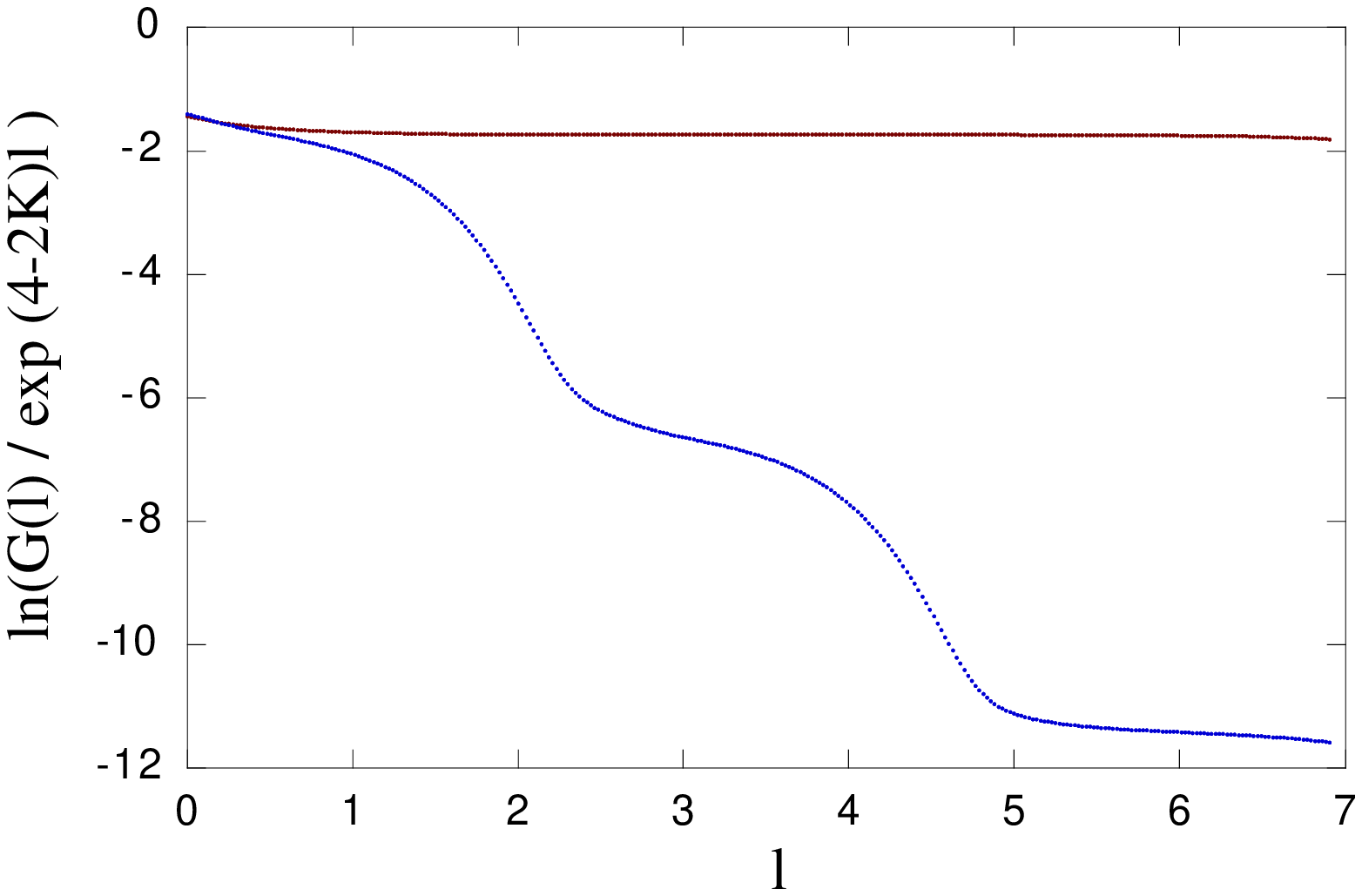}}
\caption{\label{bigpeak} Behavior of $G$ for $2k_F=2.63$ (lower curve) and
$2k_F=2.4$ (upper curve).}
\end{figure}

We now investigate the case of the Fibonacci potential. For a given maximum
renormalization length scale $l_{max}$ corresponding to
accessible physical range, different cases must be distinguished.
First, if the Fermi momentum $2k_F$ is close to a main peak
of the Fourier spectrum (\ref{TF}), then,
at long distance ({\it i.\,e.} $l\sim l_{max}$),
the flow of $K$ is controlled by this harmonic (see Fig.~\ref{bigpeak}
(upper curve)) and its behaviour is similar to the periodic one.
There is a transition at $K_c=2$, with a metallic phase for $K>2$
where the quasiperiodic potential is irrelevant.
In this phase, the system has gapless excitations whose correlations
are given by (\ref{H0bos}) with renormalized parameters.
For $K<2$, the quasiperiodicity is relevant and the
system has a charge gap given by:
\begin{equation}
\Delta \sim y_{2k_F}^{1\over 2-K(0)}
\end{equation}
Note that for $K=1$ (non-interacting case), one obtains a linear
scaling of the gap opening. Our methods allows us to recover
very simply the perturbative results \cite{Sire_pertu}  derived
by quite different methods for the particular case of non-interacting
electrons. The effect of the interactions is thus essentially to change the
scaling of the gaps and to allow a MIT at $K_c=2$ (attractive interactions).

\begin{figure}
\epsfscale=400
\centerline{\epsfbox{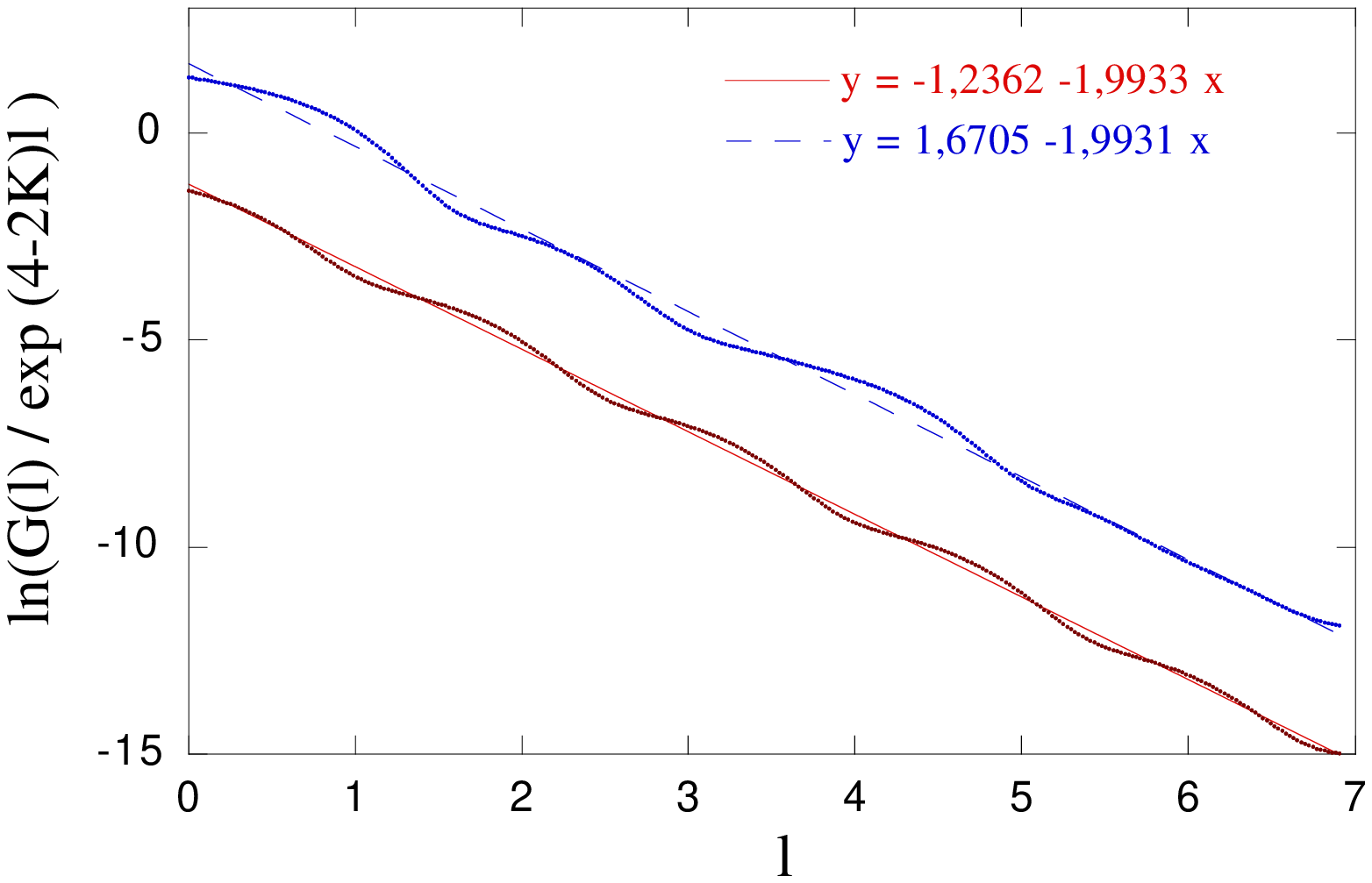}}
\caption{\label{smallpeak} Behavior of $G$
for $2k_F=\pi$ (lower curve) and $2k_F=1.95$ (upper curve).
Both cases have been offset for clarity.}
\end{figure}

A more unusual behavior is encountered when the Fermi level is far from
a dominant harmonic of the quasiperiodic potential.
Indeed, the low energy properties up to $l_{max}$
are no more dominated by the ultimate presence of a gap or not
but by the precise dependence of $G$ with the scale.
The specific feature of the quasiperiodic case is that
contrary to the disordered case, $G$ {\it a priori} depends on the Fermi level.
As shown in Fig.~\ref{smallpeak}, it is reasonable to approximate this
behavior by an exponential scaling. To know whether such a description
is asymptotically correct or not, one would need an analytical calculation
of $G$, a rather complicated task.
In this context, the flow of $K$ is given by:
\begin{equation}
{dK\over dl}=- K^2 D\, e^{(4-2K-\mu)l}\
\label{mu}
\end{equation}
where $D$ is a constant.

The position of the MIT point is then given by
$K_c = 2 - \mu/2$. In addition, the Fibonacci potential
seems to provide a unique value $\mu \simeq 2$ (see  Fig.~\ref{smallpeak}),
that leads to a transition for the non-interacting point $K_c=1$.
It would be interesting to know if this result remains true beyond
the perturbative theory, and if it is a generic
property of self-similar potentials.
Finally, note that intermediate cases can also occur (see Fig.~\ref{bigpeak}
(lower curve)) for
which $G$ can not be naively approximated by an exponential law. In this
context,
one can not simply extract a critical behaviour from the RG equations.
This deserves further investigations.

So, the quasicrystal differs from a periodic one, for
which the gap only acts for a given position of the Fermi level with $K_c=2$,
and a disordered system for which the potential is
relevant regardless of the position of the Fermi level,
but below a constant critical value $K_c=3/2$.
This important modification of $K_c$ is reminiscent of a
\emph{correlated} disorder with long range correlations in space
for which the averaged disorder potential
$\overline{\hat W^*(q)\hat W(q')}=\delta_{qq'}\Delta(q)$
is not constant.

The consequences of the scale dependence of $G$ can be directly
seen on transport quantities. In the regime where the RG
equations are valid, the conductivity can be computed using a
perturbative method. The simplest is to use the so-called memory
function formalism which is well suited to the bosonization
representation \cite{giamarchi_mott_ref}. We refer the readers to
\cite{vidal_quasiinter_long}
for the technical details and only give here the main results.
The high frequency optical conductivity and the dependence of the
resistance with respect to the system size $L=e^l$ are given by:
\begin{eqnarray}
\sigma(\omega) &\propto& M(l_\omega)/\omega^2 \\
R(L) &\propto& L M(l_L) \label{eq:rscaling}
\end{eqnarray}
where in each case the memory function $M$ is computed at a renormalized
frequency scale $l_\omega = \log(t/\omega)$ or at a renormalized
size scale $l_L = \log(L/\alpha(0))$.
$M$ is perturbatively given by \cite{giamarchi_umklapp_1d} :
\begin{equation} \label{eq:mscaling}
M(l) \propto G(l)e^{-l}
\end{equation}
This expression remains valid as long as one remains in the
perturbative regime, i.e. as long as the renormalization of $K$ in
(\ref{recy1}) remains small. This is true for all length scales in
the metallic regime $K > K_c$, where the potential $W$ is irrelevant.
For $K < K_c$ this holds until a certain length scale $L_{SC}$ corresponding
to the strong coupling limit.
Using (\ref{eq:rscaling}) and (\ref{mu}) one straightforwardly gets
the resistance for the quasiperiodic system:
\begin{equation}
R(L)\sim L^{4-2K - \mu}
\end{equation}
so that $R(L)\sim L^{4-2K}$, $R(L)\sim L^{3-2K}$ for the commensurate and
disordered cases respectively.
As foreseen, in the regime where the scattering potential is
relevant the resistance increases with the scale.
For noninteracting ($K=1$) disordered electrons we  simply recover the
Ohm's law,
which is valid below the localization length. For a quasicrystal with a dense
Fourier spectrum,
we find  a non universal power law increase of the resistance, with
an exponent depending \emph{both} on the interactions and on the position of
the Fermi level. Correspondingly, there is a power law frequency dependence of
the optical conductivity $\sigma(\omega)\sim(1/\omega)^{5-2K-\mu}$. The
properties of potential $W$ have thus a direct impact
on the increase of the resistance. This is in contrast with both the disordered
and the commensurate cases where the scaling of the resistance only depends
upon the interactions. Such a behavior and the relationship between the
conductivity
exponent and the spectrum of the quasiperiodic potential should be testable in
numerical simulations.  For the noninteracting case,
some numerical observations of a power law behavior for the
conductivity have already been reported \cite{Kubo_Goda}. However
the system sizes considered were too small to allow a precise
determination of the exponents, so that more numerical studies are
obviously needed \cite{warning_landauer}.

For $K<K_c$ the previous perturbative analysis ceases to be
valid beyond the length scale $L_{SC}$ for which $G(l_{SC}) \sim 1$.
 For the commensurate and the disorder case this defines
the correlation length respectively associated to the gap and to
the localization length.
Above $L_{SC}$ the resistance grows exponentially with the size of the
system $R(L) \sim e^{L/L_{SC}}$. For the quasiperiodic case our
analysis shows that a corresponding typical length exists as well. However its
physical interpretation and the physical behavior above $L_{SC}$ are
still to be understood. Since for a non-interacting system the
wavefunctions of a quasiperiodic system exhibit algebraic decay, a
reasonable guess is that above $L_{SC}$ the resistance keeps on increasing
as a power law but very likely with another exponent than
$4-2K-\mu$. Clearly the investigation of this strong coupling regime
deserves further studies.\\

We thank R. Mosseri, B. Dou\c cot and Cl. Aslangul for fruitful discussions.


\begin{thebibliography}{10}
\bibitem{caution_shift}
To absorb the long wavelength part of the potential $W_i$, it is necessary, as
  for disordered systems \cite{giamarchi_loc}, to perform a shift $\phi \to
  \phi + \int^x dy W(y)$ both in $H_W$ and $H_0$. At the order in perturbation
  considered here this shift has no effect \cite{vidal_quasiinter_long} and is
  omitted in the following.
\bibitem{warning_landauer}
Note that in a Landauer calculation for non-interacting electrons contact
  resistance between the leads and the system can give spurious contributions.
\end{thebibliography}
\end{document}